\documentclass[twocolumn, astrosymb, twocolappendix]{aastex631}

\usepackage{booktabs} 
\usepackage{tabularx}

\usepackage{amsmath}
\usepackage{graphicx} 
\usepackage{natbib}
\usepackage{gensymb}
\usepackage{comment}

\usepackage{float} 
\usepackage{array}

\begin{document}

\title{\textbf{LISA's view of the Galactic Halo:  forecasts for the Galactic double white dwarf population using Gaia data}}

\author[0000-0002-6394-7993]{Ann-Marsha Alexis}
\affiliation{McWilliams Center for Cosmology and Astrophysics, Department of Physics,
Carnegie Mellon University, Pittsburgh, PA 15213, USA}
\email{aalexis@andrew.cmu.edu}

\author[0000-0001-5228-6598]{Katelyn Breivik}
\affiliation{McWilliams Center for Cosmology and Astrophysics, Department of Physics,
Carnegie Mellon University, Pittsburgh, PA 15213, USA}
\email{kbreivik@andrew.cmu.edu}



\begin{abstract}

The population of close double white dwarfs (DWDs) in the Milky Way will make up the largest population of sources resolved by LISA, with a subset of the population having three-dimensional position and chirp mass measurements obtained from LISA observations. Because white dwarfs are the bulk of the Milky Way's stellar remnant population, the positions and masses of close DWDs resolved by LISA are defined by the stellar population properties that host them. Recent Gaia data has unveiled a triaxial accreted component of the Galactic stellar halo: the Gaia-Sausage-Enceladus, which contains stars that are more metal-rich than the extremely metal-poor population of stars residing in the stellar halo beyond $ 30 \, \rm{kpc} $.  In this work, we assess the size and characteristics of the population of close DWDs using an empirically motivated Galactic model which incorporates the GSE and compare to the classical Galactic model that contains only a single very metal-poor halo population. To do this, we simulate a realistic present-day Galactic DWD population and determine its gravitational wave signal in LISA 
using \texttt{LEGWORK}. We find that incorporating the metal-rich population from the GSE imprints significant differences in the chirp mass and distance distributions of resolved DWDs, but that the strength and height of the gravitational wave foreground remains unchanged.

\end{abstract}

\keywords{\href{https://astrothesaurus.org/uat/1060}{Milky Way stellar halo (1060)} --- \href{https://astrothesaurus.org/uat/1799}{White dwarf stars (1799)}  --- \href{https://astrothesaurus.org/uat/678} {Gravitational waves (678)}	
 }


\section{Introduction} \label{sec:intro}

The Laser Interferometer Space Antenna (LISA), set to launch in the 2030s, will be
sensitive to gravitational wave (GW) frequencies of $10^{-4}$ Hz to $10^{-1}$ Hz \citep{2023LRR....26....2A}. The dominant source in this range by number is the population of close double white dwarfs (DWDs) residing in the Milky Way. All of these DWDs will contribute GW emission to the LISA data stream, creating a superposition of signals called the DWD foreground  \citep[e.g.][]{Hils+1990:1990ApJ...360...75H}. Above $ \sim 3\, \rm mHz$, LISA will be able to resolve a subset of these, $\mathcal{O}(10^4)$, as individual binaries \citep{tang2024predictinggravitationalwavesignals}.

Due to their strong signal in LISA, DWDs can be complementary probes of the spatial distribution of stellar populations in the Milky Way since GW observatories are not subject to detection limits like their electromagnetic counterparts  \citep[e.g.][]{korol2019, Lamberts_2019, breivik2020constraininggalacticstructurelisa}. Studying stellar populations in the Milky Way provides an up-close picture of how galaxies form and evolve \citep{kim2026}. With astrometric data released from \textit{Gaia} DR2, which delivered stellar parallaxes, proper motions \citep{Gaia} and radial velocities from spectroscopic survey follow-ups, our understanding of Milky Way structure has been transformed, facilitating the study of the Galaxy's dynamical evolution \citep{naidu2021}.



One population, known as the \textit{Gaia-Sausage-Enceladus} (GSE), has been studied in great detail as a recent accretion event that has shaped both the Galactic thick disk and halo. The GSE was initially discovered via a population of inner halo stars with highly eccentric orbits and a coherent metallicity sequence of $[\rm{Fe/H}] < -1$ \citep{helmi2018, belokurov2018, haywood2018, myeong2018, koppelman2018, mackereth2019}.
Subsequent studies of the GSE accreted halo component characterized it as a triaxial ellipsoid \citep{iorio2019, han2022} that populates nearly half of the halo's mass \citep{naidu2020, feuillet2021, naidu2021} with an age of $\sim10\,\rm{Gyr}$ \citep{vincenzo2019, gallart2019}. 
The merger event that created the GSE is also thought to have dynamically heated the Milky Way's original disk to create the thick disk and scatter stars' orbits into a more spherically symmetric ``in situ halo" structure \citep{hayden2017ambre, dimatteo2019, belokurov2020}.

In this study, we apply new Galactic model assumptions to simulations of DWDs to investigate how the inclusion of the GSE modifies predictions for the population of DWDs that LISA will observe. This builds on previous population synthesis studies of LISA Milky Way DWDs which do not include a treatment of the GSE in their understanding of the star formation history of the Galaxy \citep[e.g.][]{nelemans2001a, ruiter2009contribution, liu2010,  yujeffery2010, nissanke2012, yujeffery2013, korol2017, kremer2017, breivik2018, Lamberts_2019, biscoveanu2023, li2023}. We discuss our methods to create Galactic populations of DWDs in Section~\ref{sec:methods}. In Section~\ref{sec:results}, we discuss our results for the Galactic population, the LISA foreground, and the resolved DWD population and compare our results to other work in Section~\ref{sec:cite}. We finish with our conclusions in Section~\ref{sec:conclusion}.

\section{Methods} \label{sec:methods}

We consider two star formation history (SFH) models in this study: a fiducial model that is representative of previous population synthesis forecasts and a new, empirically-motivated model that incorporates recent observations of the halo and thick disk stellar populations in the Milky Way. In this section, we describe the SFH assumptions for the fiducial and empirical models and the process for creating Milky Way DWD populations from COSMIC data simulated in \citet{thiele2023applying}.

\begin{table*}[ht]
    \centering
    \begin{tabular}{l c c c c c c}
        \toprule
        Component & Spatial Density & Mass ($M_{\odot}) $ & Age (Gyr) & [Fe/H] & Density Ref & Age/Metallicity Ref \\
        \midrule
        \multicolumn{7}{c}{Fiducial Model} \\
        \cmidrule(lr){1-7}
        Halo & $ \propto \hspace{0.125cm}(1+r/a_{\rm 0,halo})^{-3.5}$ & $1.4 \times 10^9 $ & 14 & -2.3 & \citealt{ruiter2010} & \citealt{robin2003synthetic} \\
        \addlinespace
        Bulge & $ \frac{\rho_{\rm {b,0}}}{(1+r'/r_{0})^{\alpha}} \exp(-(\frac{r'}{r_{\rm cut}})^2) $ & $8.9 \times 10^9 $ & 10 & 0 & \citealt{mcmillan2011mass} & \citealt{robin2003synthetic}  \\
        \addlinespace
        Thick Disk & $\Sigma_{\rm {d,0}}/(2z_{\rm d}) \exp(-\lvert z \rvert/z_{\rm d}-R/R_{\rm d})$ & $1.4 \times 10^{10}$ & 11 & -0.82 & \citealt{mcmillan2011mass} & \citealt{robin2003synthetic} \\
        \addlinespace
        Thin Disk & $\Sigma_{\rm d,0}/(2z_{\rm d}) \exp(-\lvert z \rvert/z_{\rm d}-R/R_{\rm d})$ & $4.3 \times 10^{10}$ & 1-10 & 0 & \citealt{mcmillan2011mass} & \citealt{robin2003synthetic}  \\
        \midrule
        \multicolumn{7}{c}{Empirical Model} \\
        \cmidrule(lr){1-7}
        GSE Halo & $ \propto \hspace{0.125cm} [X^2+(Y/p)^2 + (Z/q)^2]^{\alpha/2} $ 
        & $ 6 \times 10^8 $ & 10 & -1.2  & \citealt{han2022} & \citealt{naidu2020} \\
        \addlinespace
        In-situ Halo & $ \propto \hspace{0.125cm}(1+r/a_{0,\rm halo})^{-3.5}$ & $ 8 \times 10^8 $ & 14 & -0.54 & \citealt{ruiter2009contribution} & \citealt{naidu2020}  \\
        \addlinespace
        Bulge & $ \frac{\rho_{\rm b, 0}}{(1+r'/r_{0})^{\alpha}} \exp(-(\frac{r'}{r_{\rm cut}})^2)$ & $8.9 \times 10^9 $ & 10 & 0 & \citealt{mcmillan2011mass} & \citealt{robin2003synthetic} \\ 
        \addlinespace
        High-$\alpha$ Disk & $ \Sigma_{\rm d,0}/(2z_{\rm d}) \exp(-\lvert z \rvert/z_{\rm d}-R/R_{\rm d})$ & $1.4 \times 10^{10}$ & 11 & -0.54 & \citealt{mcmillan2011mass} & \citealt{naidu2020} \\
        \addlinespace
        Thin Disk & $ \Sigma_{\rm d,0}/(2{\rm z_{\rm d} }) \exp(-\lvert z \rvert/z_{\rm d} -R/R_{\rm d})$ & $4.3 \times 10^{10}$ & 1-10 & 0 & \citealt{ruiter2010} & \citealt{robin2003synthetic} \\
        \bottomrule
    \end{tabular}
    \caption{The spatial density, mass, age, and metallicity of the Milky Way components for each model, sorted by the component mass. The scale radius in the fiducial halo, $a_{\rm 0,halo}$, is $ 3.5 \ \rm{kpc} $ . The density profile of the bulge is given with values of  $\alpha = 1.8$, $r_{0} = 0.075\,\rm{kpc}$, $ r_{\rm cut} = 2.1 \ \rm{kpc} $ and the axial ratio of $q = 0.5$. In the disk models, the scale heights are given by $z_{\rm d,thin} = 300 \ \rm{pc}$ and $z_{\rm d,thick} = 900 \ \rm pc$. The scale lengths, $R_{\rm d}$, for the thin and thick discs are $ 2.6 $ and $ 3.6 \ \rm{kpc} $ for the thin and thick disks, respectively. The surface density, $\Sigma_{\rm {d,0}}$, is given by $816.6$ for the thin disk  and $209.5$ for the thick disk. For the GSE halo, $\alpha$ is determined in three different regions characterized by the breaks in radius, with $\alpha_1$ being $1.70$, $\alpha_2$ being $3.09$ and $\alpha_3$ being $4.58$; the values p and q are $0.81$ and $0.73$. }
    \label{table:1}
    
\end{table*}
\subsection{Milky Way SFH assumptions}\label{sec:SFH}


A summary of our model assumptions is given in Table~\ref{table:1}, where each Galactic component is assigned an analytic distribution function and single metallicity. The spatial density distributions, masses, agees, and metallicities are listed for each component and each model. The references which inform each choice are also listed for the spatial density and age/metallicity. We note that this approach differs from studies which use cosmological zoom in simulations like the {\textbf{m12i}} galaxy from the Latte Suite of the FIRE-2 simulations \citep[e.g.][]{Lamberts_2019, thiele2023applying, tang2024predictinggravitationalwavesignals}. 

In our fiducial model, we follow the SFH and spatial distributions described in \citealt{mcmillan2011mass}, \citealt{robin2003synthetic} and \citealt{ruiter2009contribution}. All metallicities are assigned relative to the solar abundance pattern such that $\rm{[Fe/H]} = \log_{10}(Z/Z_\odot)$ and $Z_\odot = 0.02$ and are obtained from \citealt{robin2003synthetic}. We assume that the thin disk has a constant, solar metallicity star formation over the past $10\,\rm{Gyr}$ and that the thick disk formed through a $1\,\rm{Gyr}$ burst of star formation $11\,\rm{Gyr}$ in the past with $15\%$ solar metallicity. We further assume that the bulge formed from a $1 \, \rm{Gyr}$ burst of solar-metallicity star formation $10\,\rm{Gyr}$ in the past. Finally, we assume that the halo formed as a $1 \, \rm{Gyr}$ burst of $0.5\%$ solar metallicity, 
$14\,\rm{Gyr}$ in the past. The masses for the thin disk, thick disk, and bulge follow \citealt{mcmillan2011mass} and the mass of the halo distribution follows \citealt{ruiter2009contribution}. 

In our empirical model, we apply the same SFH assumptions for the thin disk and bulge, but apply empirically motivated assumptions for the formation of the halo and thick disk, mediated via the GSE accretion event as described in \citealt{naidu2020}.
We split the halo into two components: the GSE halo and the in-situ halo.
We use the same total halo mass from \citealt{robin2003synthetic} but attribute $42\%$ of the halo mass to the GSE halo following \citet{han2022} and the remaining $58\%$ to the in-situ halo. We assume that the GSE member stars experienced a $1 \, \rm{Gyr} $ period of star formation $10 \,\rm{Gyr}$ ago with a metallicity of $\rm{[Fe/H]}=-1.2$ and that the in-situ halo component formed from a $1\,\rm{Gyr}$ burst of star formation $14 \, \rm{Gyr} $ in the past with a metallicity of $\rm{[Fe/H]}=-0.54$. 
We assume that the thick disk formed from a `high-$\alpha$' $1 \,\rm{Gyr}$ burst of star formation with $\rm{[Fe/H]} = -0.54$ $11\,\rm{Gyr}$ in the past that was truncated by the GSE accretion event. In the rest of this paper, we use the term empirical halo to describe the combined GSE and in-situ halo components.

\subsection{Data Retrieval}
We use the publicly available data from the fiducial binary evolution model described in \citealt{thiele2023applying} which applies the metallicity-dependent close binary fraction from \citealt{moe2019}. 
These populations were simulated using the \texttt{COSMIC} binary population synthesis suite \citep{breivik2020cosmic} for a grid of fifteen metallicities uniformly spaced as  $ -2.3 <  \rm [Fe/H] =< 0.18$.
Similar to \citealt{thiele2023applying}, we consider four types of DWD binaries: helium white dwarf pairs (HeHe), carbon-oxygen and helium DWDs (HeCO), carbon-oxygen white dwarf pairs (COCO) and oxygen-neon white dwarfs with helium, carbon-oxygen, or oxygen-neon white dwarf companions (ONeX). 

For each DWD type, we select binaries with a separation of less than $1000\,R_{\odot}$ to ensure that we only consider DWDs that could evolve to become LISA sources at present day. 

For each Milky Way component in our fiducial and empirical models, we select the simulated population that matches most closely to the component metallicity for each DWD type, under the assumption that all stars in each component form with the same metallicity. The metallicity values of each component are described in Table~\ref{table:1}. 

\subsection{Creating Milky Way populations}

For each Galactic component, we compute the formation efficiency, $\eta_{\rm{form}}$, or the number of DWDs formed per unit stellar mass as

\begin{equation}
    \eta_{\rm form} = \frac{N_{\rm{DWD,sim}}}{M_{\rm sim}},
\end{equation}
\noindent where $N_{\rm{DWD,sim}}$ is the number of simulated DWDs with separations below $1000\,R_{\odot}$ and $M_{\rm sim}$ is the total stellar mass required to produce the simulated DWD population.
The number of DWDs in the component is then calculated by scaling the formation efficiency by the mass of the component, $M_{\rm{component}}$, as 
\begin{equation}
    N_{\rm DWD,\,component} = \eta_{\rm form} \times M_{\rm{component}}.
\end{equation}

In order to predict the rates and binary properties of the present-day population of Milky Way DWDs, we sample $N_{\rm DWD}$ DWDs with replacement from the simulated data and assign ages and positions to each sampled DWD binary according to the SFH of their respective Galactic component. The ages are assigned such that $t_{\rm{age}}$ represents the present-day age from the Zero Age Main Sequence formation time of each DWD progenitor binary. We then compute the GW evolution time as the time that has elapsed since the formation of the simulated DWD ($t_{\rm{DWD}}$) as
\begin{equation}
    t_{\rm{GW\,evol}} = t_{\rm{age}} - t_{\rm{DWD}}.
\end{equation}

To construct a present-day DWD population,
we filter any systems for which $t_{\rm{age}}<t_{\rm{DWD}}$ since their progenitor binaries do not have time to evolve to produce a DWD by the present day. For DWDs 
which form before the present day, we evolve their orbits due to GW emission using \texttt{LEGWORK} \citep{LEGWORK_joss,LEGWORK_apjs,legwork_15579534} following \citet{Peters+1964:1964PhRv..136.1224P} and remove any DWDs that fill their Roche lobes before the present day.
 
We finally require that the DWDs have present-day orbital frequencies such that their GW frequencies are within the LISA frequency band between $10^{-4}\,\rm{Hz} < f_{\rm{GW}} < 10^{-1}\,\rm{Hz}$ (where $f_{\rm{GW}} = 2f_{\rm{orb}}$). 
For DWDs radiating GWs in the LISA frequency band, we assign positions following the spatial density profiles described in Table~\ref{table:2}.

\subsection{LISA Detectability}

We use \texttt{LEGWORK}, which follows the derivations in \cite{Flanagan&Hughes} and uses the LISA noise power spectral density (PSD) in \cite{Robson+2019:2019CQGra..36j5011R}, to calculate the SNR of the DWDs in the LISA band assuming an observation duration of $T_{\rm obs}=4\,\rm{yr}$. \texttt{LEGWORK} considers the frequency change for quadrupole GW emission to lowest-order in the post-Newtonian expansion, defined as

\begin{equation}
    \dot{f_{\rm 2}} = \frac{96}{5\pi}\frac{(G\mathcal{M}_{\rm c})^{5/3}}{c^5}(2\pi f_{\rm{orb}})^{11/3}.
\end{equation}

\noindent for circular orbits, where $\mathcal{M}_{\rm{c}}$ is the chirp mass of the binary. 

For sources whose orbital frequency evolves appreciably over the observation duration, the SNR is calculated as

\begin{equation}
\langle\rho \rangle ^2_{\rm circ,evol} =  \int_{2f_{\rm{orb,0}}}^{2f_{\rm{orb,1}}}{df_{\rm{orb}} \frac{h^2_{\rm c,2}}{f_{\rm{orb}}^2S_{\rm n}(2f_{\rm{orb}})}}.
\end{equation}

\noindent where $S_{\rm n}$ is the LISA sensitivity curve defined in \citealt{Robson+2019:2019CQGra..36j5011R} and $h_{\rm c,2}$ is the characteristic strain.\footnote{We note that an increased observation duration will increase the number of detected DWDs but will not significantly alter any of our results discussed in Section~\ref{sec:results}.} For circular orbits, the characteristic strain is

\begin{equation}
    h_{\rm c, 2}^2 = \frac{2^{2/3}}{3\pi^{3/4}}\frac{(G\mathcal{M}_{\rm c})^{5/3}}{c^3 D^2_{\rm L}}\frac{1} {f_{\rm{orb}}^{1/3}}.
\end{equation}

\noindent Here, 
$D_{\rm L}$ is the luminosity distance which we take to be the distance between each DWD and the Sun based on our position assignments.

For stationary sources that do not evolve appreciably during the observation, the SNR simplifies to

\begin{equation}
\rho^2_{\rm circ,stat} \approx \frac{1}{4}\frac{h^2_{0,2}T_{\rm obs}}{S_{\rm n}(2f_{\rm{orb}})} = \frac{1}{4}\frac{PSD}{S_{\rm{n}}(2f_{\rm{orb}})}, 
\end{equation}

\noindent where $h_{0,2}$ is the strain amplitude of the source 
for circular orbits {and defined as, 

\begin{equation}
h_{0,2}^2 = \frac{ 2^{22/3} } {5} \frac{ (G \mathcal{M}_{\rm c} ) ^{10/3} } {c^8 D_{\rm L}^2}\left(\pi f_{\rm orb} \right)^{4/3} \,.
\end{equation}




\noindent We can then approximate a stationary binary's SNR with the amplitude spectral density defined as 

\begin{equation}
ASD = \sqrt{PSD} =  h_{0,2}\sqrt{T_{\rm obs}},
\end{equation}

\noindent such that

\begin{equation}
    \rho_{\rm circ,stat} = \frac{1}{2}\frac{h_{0,2} \sqrt{T_{\rm obs}}}{\sqrt{S_{\rm n}(2f_{\rm{orb}})}} = \frac{1}{2}\frac{ASD}{\sqrt{S_{\rm{n}}(2f_{\rm{orb}})}}.
\end{equation}




We assume that any DWD with $ \langle\rho \rangle = \rm{SNR}>7$ is resolved as an individual source and contributes to the LISA-resolved population.

\begin{table*}[ht]
        \centering
        
\begin{tabular}{lrrrrrrr}

\toprule
Components & Type & $N_{\rm total}$ & $N_{\rm merged}$ & $N_{\rm detached}$ & $N_{\rm f_{GW} > 1e-4 \, Hz}$  & $N_{\rm SNR >7}$ \\

\midrule

Thin Disk & HeHe  & $1.07\times10^7$ & $1.01\times10^7$ & $6.43\times10^5$ & $5.71\times10^5$ & $6625$ \\
& HeCO & $8.13\times10^7$ & $5.78\times10^7$ & $2.35\times10^7$ & $6.95\times10^6$ & $20556$ \\
& COCO  & $4.83\times10^7$ & $1.50\times10^7$ & $3.33\times10^7$ & $1.13\times10^6$ & $3367$ \\
& ONeX & $1.06\times10^7$ & $5.15\times10^6$ & $5.41\times10^6$ & $3.05\times10^5$  & $1013$ \\
\hline
Total & & $1.51\times10^8$ & $8.81\times10^7$ & $6.28\times10^7$ & $8.95\times10^6$  & $31561$ \\

\hline
\hline

Bulge & HeHe & $5.39\times10^6$ & $4.98\times10^6$ & $4.12\times10^5$ & $3.36\times10^5$ & $1397$ \\
& HeCO  &  $2.71\times10^7$ & $1.88\times10^7$ & $8.35\times10^6$ & $1.70\times10^6$  & $2016$ \\
& COCO & $1.19\times10^7$ & $3.72\times10^6$ & $8.23\times10^6$ & $1.72\times10^5$  & $130$ \\
& ONeX & $2.47\times10^6$ & $1.26\times10^6$ & $1.20\times10^6$ & $1.29\times10^4$  & $9$ \\
\hline
Total  & & $4.69\times10^7$ & $2.87\times10^7$ & $1.82\times10^7$ & $2.22\times10^6$  & $3552$ \\

\hline
\hline
\hline
\vspace{10pt} \\

\hline
\hline
Fiducial Halo & HeHe  & $4.42\times10^6$  & $3.84\times10^6$ & $5.79\times10^6$ & $1.06\times10^5$ & $270$ \\

& HeCO  & $8.21\times10^6$  & $5.48\times10^6$ & $2.73\times10^6$ & $2.99\times10^5$ & $247$ \\

& COCO  & $3.64\times10^6$ & $5.65\times10^5$ & $3.07\times10^6$ & $2.47\times10^4$ & $29$ \\

& ONeX  & $6.79\times10^5$ & $2.91\times10^5$ & $3.88\times10^5$ & $4.88\times10^3$ & $5$ \\

\hline
Total & & $1.69\times10^7$ & $1.02\times10^7$ & $6.77\times10^6$ & $4.35\times10^5$  & $551$ \\

\hline
\hline

Fiducial Thick Disk & HeHe & $4.49\times10^7$ & $4.02\times10^7$ & $4.65\times10^6$ & 
$1.91\times10^6$ & $4437$ \\

& HeCO  & $6.86\times10^7$ & $4.74\times10^7$ & $2.12\times10^7$ & $3.44\times10^6$ & $3434$ \\

& COCO  &  $2.84\times10^7$ & $1.27\times10^7$ & $1.57\times10^7$ & $5.18\times10^4$ & $48$ \\

& ONeX  & $6.06\times10^6$ & $3.61\times10^6$ & $2.45\times10^6$ & $2.90\times10^4$ & $29$ \\

\hline
Total & & $1.48\times10^8$ & $1.04\times10^8$ & $4.40\times10^7$ & $5.43\times10^6$ & $7948$ \\

\hline
\hline

\textbf{Fiducial Galactic Total} & & $3.63\times10^8$ & $2.31\times10^8$ & $1.32\times10^8$ & $1.70\times10^7$ & $43612$ &  \\
   
\hline
\hline
\vspace{10pt} \\

\hline
\hline
In Situ Halo & HeHe  & $9.55\times10^5$ & $8.80\times10^5$ & $7.49\times10^4$ & $3.91\times10^4$ & $100$ \\

& HeCO & $3.51\times10^6$ & $2.38\times10^6$ & $1.12\times10^6$ & $1.51\times10^5$ & $147$ \\

& COCO & $1.48\times10^6$ & $3.51\times10^5$ & $1.13\times10^6$ & $1.17\times10^4$  & $4$ \\

& ONeX  & $3.11\times10^5$ & $1.49\times10^5$ & $1.63\times10^5$ & $1.67\times10^3$ & $0$ \\

\hline

Total & & $6.26\times10^6$ & $3.76\times10^6$ & $2.49\times10^6$ & $2.03\times10^5$  & $251$ \\

\hline
\hline

GSE Halo & HeHe & $1.75\times10^6$ & $1.64\times10^6$ & $1.09\times10^5$& $5.05\times10^4$ & $104$ \\

& HeCO  & $3.02\times10^6$ & $1.98\times10^6$ & $1.03\times10^6$ & $1.67\times10^5$  & $112$ \\

& COCO  & $1.26\times10^6$ & $4.03\times10^5$ & $8.58\times10^5$ & $5.80\times10^3$  & $2$ \\

& ONeX  & $2.95\times10^5$ & $1.59\times10^5$ & $1.36\times10^5$ & $8.65\times10^2$  & $0$ \\

\hline
Total & & $6.33\times10^6$ & $4.19\times10^6$ & $2.14\times10^6$ & $2.24\times10^5$ & $218$ \\

\hline
\hline

Empirical Thick Disk & HeHe & $1.41\times10^7$ & $1.34\times10^7$ & $6.78\times10^5$ & $5.08\times10^5$ & $2347$ \\

& HeCO & $5.74\times10^7$ & $3.90\times10^7$ & $1.84\times10^7$ & $3.16\times10^6$ & $3575$ \\

& COCO & $2.39\times10^7$ & $5.94\times10^6$ & $1.80\times10^7$ & $2.88\times10^5$  & $290$ \\

& ONeX & $5.50\times10^6$ & $2.61\times10^6$ & $2.89\times10^6$ & $4.48\times10^4$  & $45$ \\

\hline

Total & & $1.01\times10^8$ 
& $6.09\times10^7$ & $4.00\times10^7$ & $4.00\times10^6$ & $6257$ \\

\hline
\hline

\textbf{Empirical Galactic Total} & & $3.11\times10^8$ & $1.86\times10^8$ & $1.26\times10^8$ & $1.56\times10^7$ & $41730$ &  \\
\hline

\bottomrule

\end{tabular}

    \caption{The total number of DWDs, the DWDs that have merged, remain detached with separations below $ 1000 R_{\odot}$, radiate GWs in the LISA band, and are resolved with $\rm{SNR} > 7$. The thin disk and bulge is accounted for in the total for both the fiducial Galactic total and the empirical Galactic total.}    
    \label{table:2}
    
\end{table*}


\section{Results}\label{sec:results}

In the following section, we show how applying our new empirical model for the Milky Way SFH affects the DWD population that LISA may observe. We first discuss how the changes to metallicity in our empirical model affects the formation efficiency. We then compare our predictions between the fiducial and empirical models for the DWD foreground and resolved population. Since our empirical model contains updates to the thick disk and halo, we describe these populations in detail.



\subsection{The DWD formation efficiency}
\label{sec:formation-efficiency}

\begin{figure}[htbp] 
\includegraphics[width=\columnwidth]{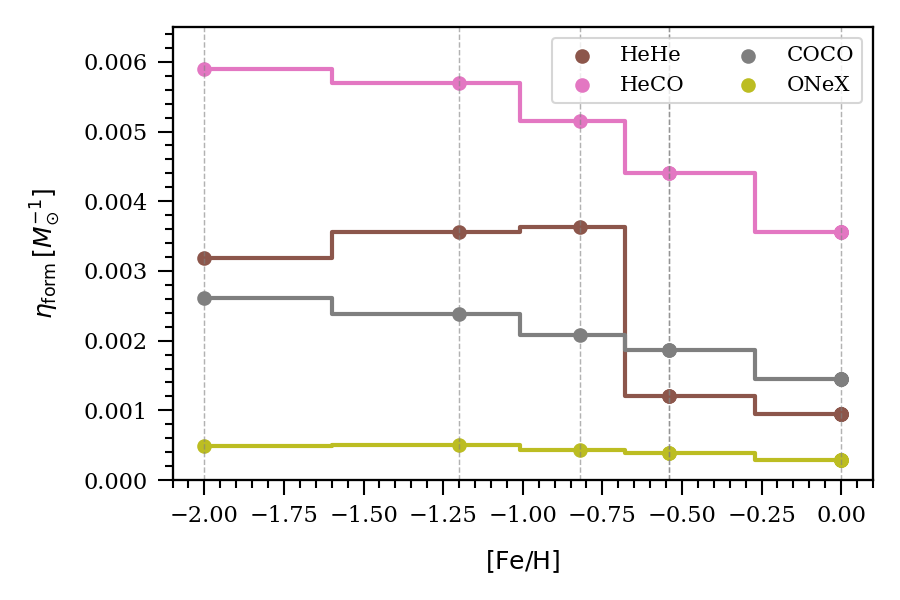}
\caption{The formation efficiency from the fiducial model of \cite{thiele2023applying} for each DWD type: HeHe (brown), HeCO (pink), COCO (grey), and ONeX (olive). Each dot and vertical dashed line corresponds to a metallicity assigned to a Galactic component in our simulations. The components in order of increasing metallicity are the fiducial halo, the GSE halo, the fiducial thick disk, the high-$\alpha$ thick disk/in-situ halo, and the bulge/thin disk. The formation efficiency, $\eta_{\rm form}$, decreases with increasing metallicity such that our empirical Galactic model has lower formation efficiencies on average when compared to the fiducial Galactic model.}
\label{fig:formation efficiencies}
\end{figure}

 The formation efficiency of each DWD type for the metallicities represented in our Galactic components is shown in Figure \ref{fig:formation efficiencies}. There is a general trend of decreasing formation efficiency with increasing metallicity where He-hosting binaries show the largest reductions. 
 One reason for the reduction in formation efficiency is that the close binary fraction decreases with increasing metallicity \citep{moe2019, mazzola2020}. In addition to the reduced binary fraction, DWDs form with lower formation efficiencies primarily due to binary interactions between their stellar progenitors, except in the case of ONe WD progenitors which have reduced formation efficiencies because they undergo enhanced line-driven wind mass loss rates. We refer the reader to \citet{thiele2023applying} for a detailed discussion of how binary interactions alter the formation of DWDs as a function of metallicity. 
 
 The general trend of decreasing formation efficiency is most pronounced in He-hosting DWDs between $ \rm [Fe/H] $ of $ -0.82$ and $-0.54$, corresponding to our fiducial and empirical thick disk assumptions, respectively.
 For HeHe DWDs, the decrease in formation efficiency does not apply to the GSE halo since the decrease occurs at metallicites higher than $ \rm [Fe/H] = -0.75 $. In this case, the GSE halo actually has a slightly increased formation efficiency for HeHe DWDs, at the higher metallicity of $\rm [Fe/H] = -1.2 $ relative to the fiducial halo of $\rm [Fe/H] = -2.3 $. The formation efficiency decreases for all other DWD types in the GSE halo. Moreover, relative to the fiducial halo, the in situ halo experiences a drop in HeHe DWD formation efficiency at its metallicity of -0.54. The overall increase in metallicity for our empirical halo model thus still results in a decrease in the formation efficiency broadly. 

\subsection{ Galactic DWD population statistics}
\label{sec:galactic-pop-numbers}

\subsubsection{Detached DWDs}
Changes to the age and metallicity between our fiducial and empirical Galaxy models result in slightly fewer total DWDs in the empirical model. We show the counts for each DWD type and Galactic component at each stage of filtering described in Section~\ref{sec:methods} in Table~\ref{table:2}. The counts for the thin disk and bulge are shown separately from the halo and thick disk since we apply the same SFH assumptions for these components in both the fiducial and empirical models. For each Galaxy model, the total number of detached DWDs with separations below $1000\,R_{\odot}$ is of order $\mathcal{O}(10^8)$ and differences between models are of order $\mathcal{O}(10^7)$. These differences are the result of modifications applied in our thick disk and halo SFH model assumptions, so we discuss these effects in detail below.

Since the thick disk constitutes more than a third of the Milky Way's mass, updating assumptions for its formation has the strongest impact in the number of detached DWDs. When adopting the higher metallicity empirical model, we find that the number of DWDs in the population decreases by $4.7\rm{e}7$ systems. In this case the drop is dominated by the decrease in HeHe DWDs which reduces by a factor of $\sim3$} and the number of HeCO, COCO and ONeX DWDs decreases by only a factor of $1.1-1.2$ when adopting the empirical thick disk assumptions, as determined by the corresponding drop in formation efficiency.

Updating the age and metallicity of the halo also results in a decreased production of DWDs by $4.3\rm{e}6$ systems. For the halo, this is not only because of the higher metallicites in the empirical halo relative to the fiducial halo, but also due to modifications to the halo's age owing to the inclusion of the GSE in the empircal model. Although the HeHe DWD formation efficiency is larger at the GSE halo's metallicity, the younger age assigned to the GSE halo leads to a reduction in the total number of present-day DWDs because a non-negligible fraction of the progenitor binaries have DWD  formation times longer than the age of the GSE halo.

\subsubsection{DWD mergers}
More than half of the DWDs that form in our Galactic population merge by the present day. Of the less than 50\% of the population that remain as detached DWDs, 12\% are closely-orbiting with present-day frequencies that will be in the LISA band. For a fixed binary evolution model, assumptions for the age and metallicity can modify the number of mergers, which also modifies the size of the detached DWD population and thus the number of LISA-resolved sources. For a more detailed discussion of the DWD merger population, we refer the reader to \cite{Kremer+2026:2026arXiv260505308K}.

Since only the metallicity is changed between the fiducial and empirical thick disk models, we expect that the formation efficiency and DWD formation properties drive differences between their merger rates. Due to the increased metallicity in the empirical thick disk, the formation efficiency of DWDs in the empirical thick disk is lower for all DWD types when compared to the fiducial thick disk, however the decrease in the merger rate does not match the fractional decrease in the formation efficiency. The separation distribution of the DWD population at formation must therefore play a significant role. For example, there are fewer He-hosting DWDs that form in the higher metallicity empirical thick disk, but the number of mergers is comparable, suggesting that the separation distribution must be smaller at DWD formation. Higher-metallicity progenitors initiate CE as more massive stars and are thus more likely to merge rather than produce a DWD. In this case, the merger rate more closely follows the formation efficiency, such that the rate of mergers decreases by $20\%$ for the COCO DWDs and $12\%$ for the ONeX DWDs.

There is a general increase in the rate of mergers for all DWD types in the in situ and GSE halo sub-components relative to the fiducial halo. The merger rate of He-hosting DWDs in the empirical is comparable to the merger rate of He-hosting DWDs in the fiducial halo within a range of $10\%$, despite the reduction in formation efficiency. As with the thick disk, this suggests that the separation distribution of He-hosting DWDs in the empirical halo must be smaller compared to He-hosting DWDs in the fiducial halo. Further, the merger rate for the COCO DWD increases by up to $15\%$ in the empirical halo by $5-11\%$ for the ONeX DWD type. We can conclude that at the higher metallicities of the empirical halo and the younger age of the GSE halo, COCO and ONeX DWDs also form with closer separations than at the lower metallicity of the fiducial halo.

\subsubsection{DWDs with frequencies in the LISA band}
For DWDs with frequencies in the LISA band, the difference between the fiducial and empirical models is small. The total number of LISA DWDs decreases by less than $10\%$ while the total number of resolved DWDs decreases by less than $5\%$. We discuss how this reduction impacts the Galactic foreground and resolved population in the following sections.

\begin{figure}[htbp] 
\includegraphics[width=\columnwidth]{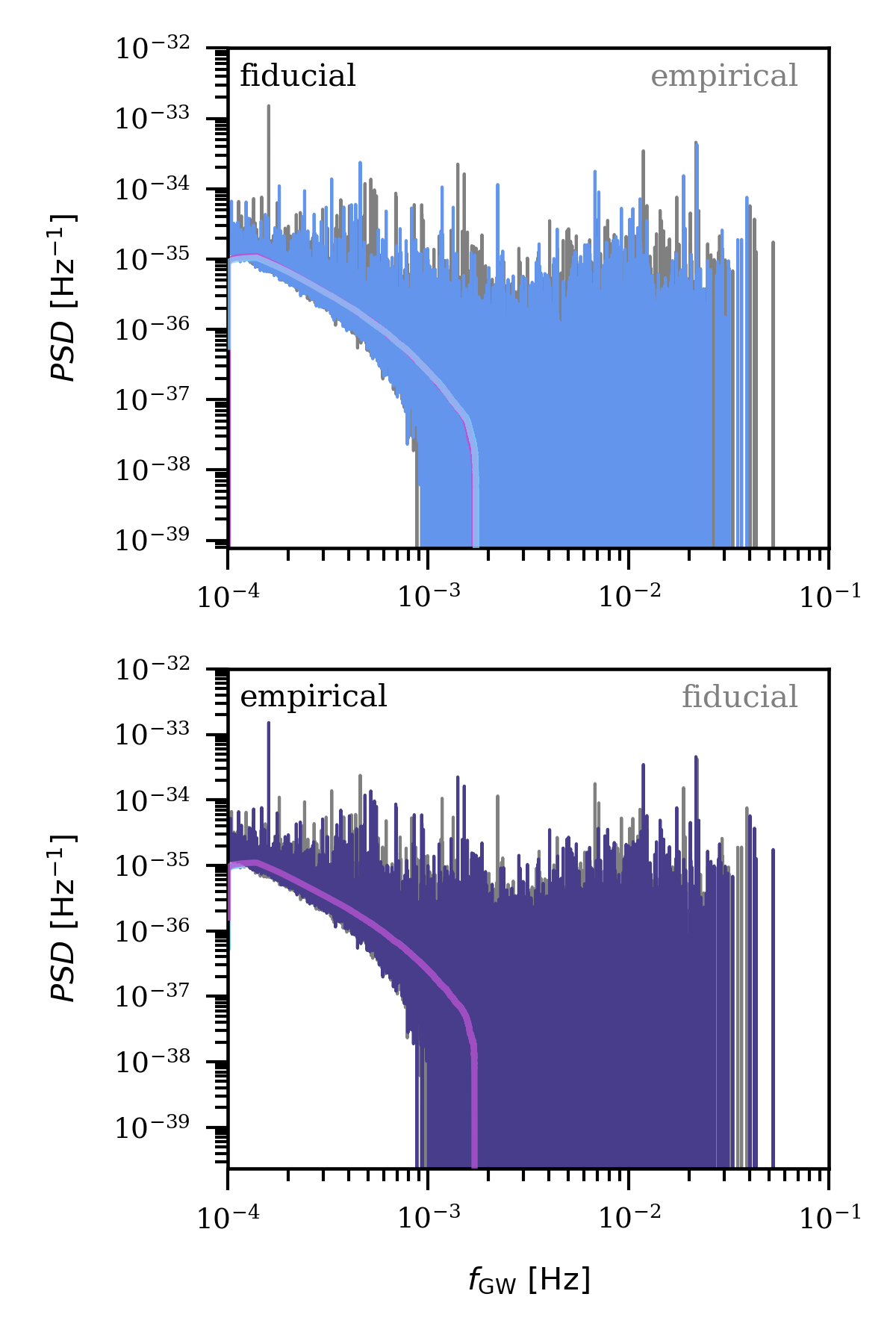}
\caption{The PSD as a function of frequency for our two Galaxy models. We approximate the irreducible foreground with a rolling median that has a window of 10,000 frequency bins (light colors). For each panel, we show a comparison of the alternate model in grey and the alternate irreducible foreground. There are only slight differences between the PSDs while the irreducible foregrounds are indistinguishable.} 
\label{fig:psd}
\end{figure} 

\subsection{The LISA DWD foreground}
\label{sec:lisa-pop}

Figure \ref{fig:psd} shows the PSD as a function of GW frequency for the fiducial (top) and empirical (bottom) Galaxy model. The alternate Galaxy model is plotted in grey for comparison in each panel while the colored lines show the PSD. 
In each panel, the irreducible foreground of both signals is approximated with a running median with $10000$ frequency bins and is plotted in the lighter blue and purple colors corresponding to the fiducial and empirical models respectively. Ultimately, the strength and the shape of the foreground and PSD remain qualitatively unchanged when adopting the empirical Galaxy model. Thus, our models suggest that the inclusion of the GSE does not alter the LISA foreground significantly. For comparison, LISA’s strain amplitude is modulated over its orbit by a factor of 1.5 \citep{cornish2003}, while our irreducible foregrounds are nearly indistinguishable.

\subsection{Resolved population and LISA observables}
\label{sec:resolved-pop}

We find that LISA will be able to detect of order $\mathcal{O}(10^4)$ GW sources with SNR $>$ 7 in each of our models,
with a decrease of less than 5\% in the number of resolved sources between models (Table~\ref{table:2}). The number of resolved sources, however, does not necessarily encapsulate the impact of our empirical model assumptions on the binary properties and spatial distribution of LISA-resolved DWDs in each Galactic component.

In the following subsections, we highlight {modifications to the LISA observables in the thick disk and halo to show how the empirical model influences predictions for the resolved DWD population. We do not discuss the thin disk or bulge since these Galactic components remain unchanged in our fiducial and empirical models. 

\subsubsection{Thick Disk}
\label{sec:thick-disk}
\begin{figure*}[htbp] 
\centering
\includegraphics[width=0.75\textwidth]{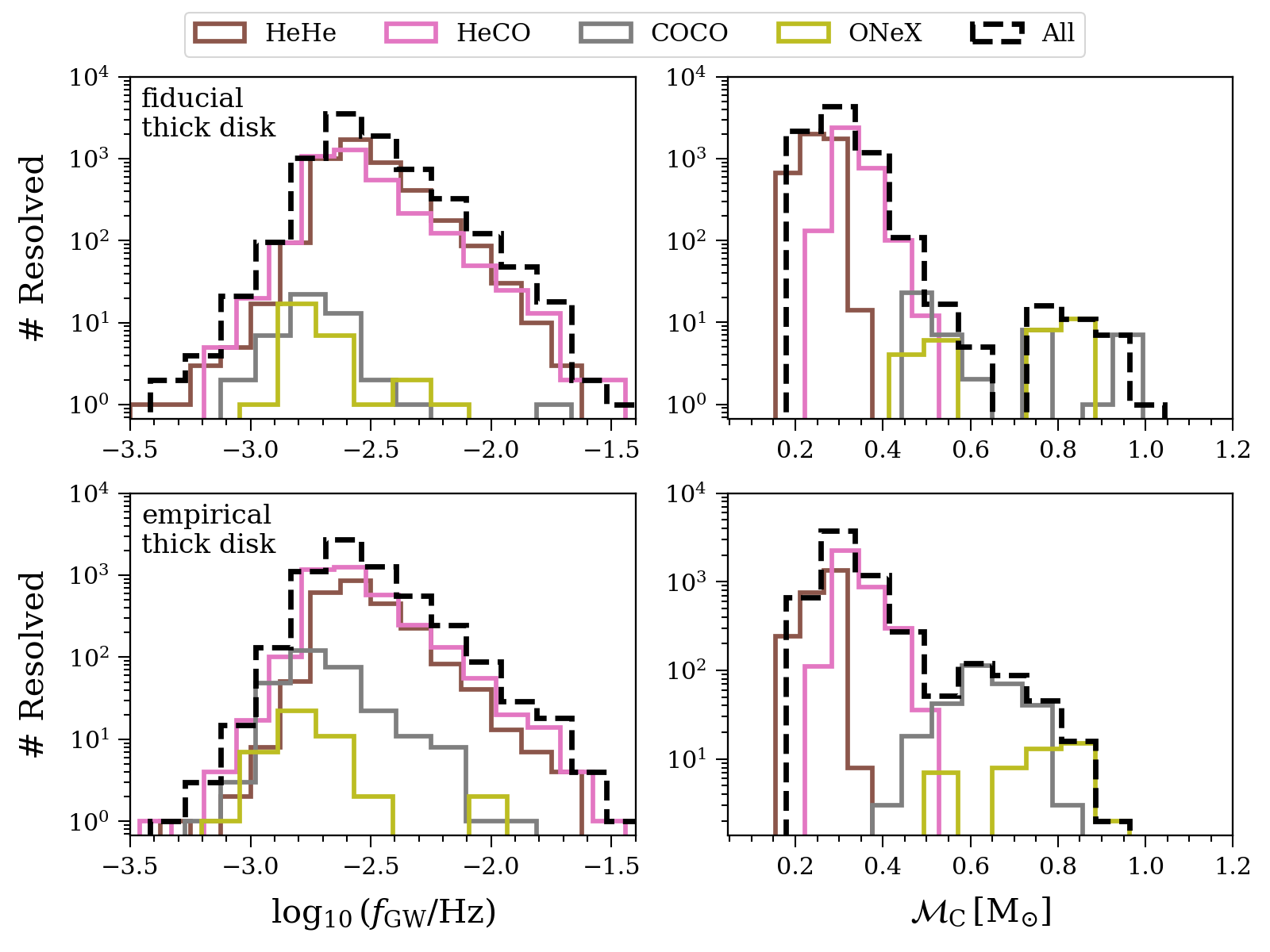}
\caption{The distribution of LISA observables of the resolved DWD LISA population for the fiducial and empirical thick disk or high-$\alpha$ thick disk. Each color represents a different DWD type (HeHe, HeCO, COCO, ONeX). The brown, bright pink, grey and olive colors correspond to DWD types HeHe, HeCO, COCO and ONeX. The dashed black line represents all the DWD types in the component. LISA will resolve more COCO DWDs within the range of 0.6-0.8 $M_{\odot}$ in the empirical thick disk.}
\label{fig:DWD LISA observables for the thick disk}
\end{figure*}

Figure~\ref {fig:DWD LISA observables for the thick disk}  shows the GW  frequency and chirp mass distributions for each Galaxy model where each DWD type is indicated with a different color. 
We find that the combined frequency distribution for all DWDs remains qualitatively unchanged between the fiducial and empirical models. However, the COCO population in the empirical thick disk extends to higher frequencies and the HeHe population contains nearly double the number of DWDs due to the increased metallicity in the empirical model.

The COCO chirp mass distribution of resolved DWDs is more populated in the empirical thick disk between $0.6-0.8 M_{\odot}$ than in the fiducial thick disk population. This population exists at higher metallicities because the DWDs form from two CE events that then lead to shorter-period DWDs at formation when compared to the lower metallicity population which forms with longer periods. For the lower metallicity fiducial thick disk, the DWDs with chirp masses between $0.6-0.8 M_{\odot}$ exist, but have lower GW frequencies since they are at wider separations such that they are not resolvable by LISA.

\subsubsection{Halo}
\label{sec:halo}

Figure \ref{fig:LISA observables for the halo} shows the distribution of metallicity, frequency, chirp mass and distance for the resolved DWDs in our fiducial halo (top) and empirical halo (bottom row). The leftmost panel shows the frequency distribution for our two halo models, where as in the thick disk, the combined frequency distribution between models is qualitatively unchanged. The middle panel shows the distance distributions, where we highlight that the distance of the resolved empirical halo DWDs extends beyond 40 kpc due to the addition of the GSE. We find that these far-away DWDs are of type HeHe and HeCO which form at higher frequencies that are thus more easily resolved at farther distances. Lastly, the rightmost panel shows the chirp mass distributions for each Galactic model where there is a reduction in the population of COCO and ONeX DWDs with chirp masses above $0.7\,M_{\odot}$. 

The chirp mass distribution of changes between the fiducial and empirical halo due to increased metallicities in the empirical halo combined with the younger GSE component age. At higher metallicities, the COCO progenitor population experiences an increased rate of CE mergers relative to the low metallicity fiducial halo population, thus reducing the overall formation rate. This compounds with the GSE having a younger age, and thus less time for the COCO DWDs that do form to evolve toward higher frequencies. 
Finally, we note that there are no resolved ONe-hosting DWDs in the empirical model. This is due, in part, to their formation being inhibited due to stronger stellar winds at higher metallicities. In addition, ONe-hosting DWDs form with wider orbits that when compounded with larger distances reduce their ability to be resolved. Therefore our models suggest that the LISA-resolved sources in the halo, especially those with distances extending beyond the fiducial halo, will be predominantly He-hosting DWDs.

\begin{figure*}[htbp] 
\includegraphics[width=\textwidth]{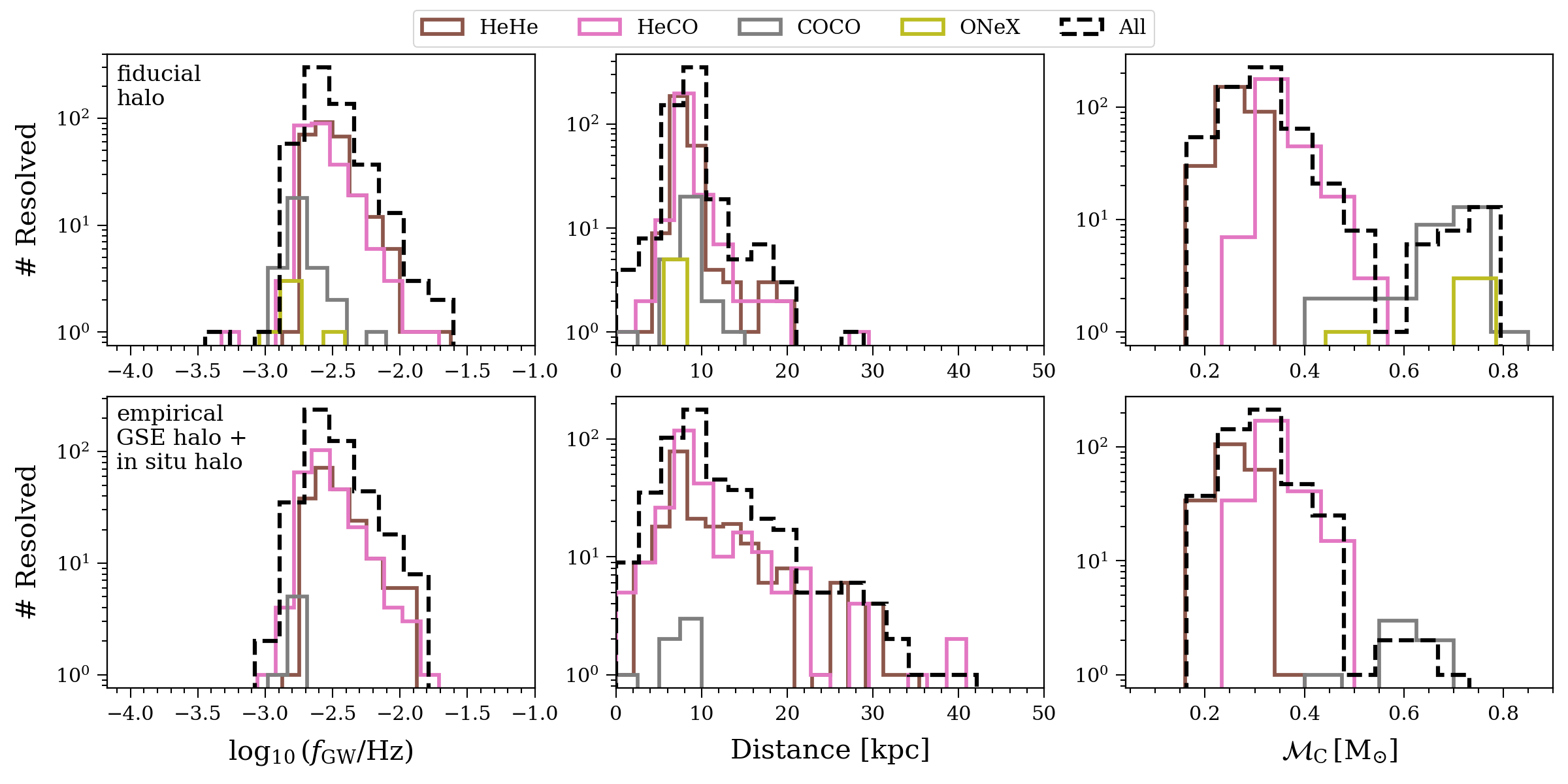}
\caption{ The frequency, chirp mass, and distance distribution of the LISA-resolved DWD population for the fiducial and empirical halo. Each color depicts a different DWD type: HeHe (brown), HeCO (pink), COCO (grey), ONeX (olive). The dashed line shows the distributions for all the DWD types for the select component. Higher frequency sources like the HeHe and HeCO DWD types are resolved at further distances, as can be seen in the empirical halo. The younger GSE halo component removes COCO DWDs with chirp masses above $ 0.7 M_{\odot}$ when compared to the fiducial model. Therefore, we predict that LISA will not detect many sources above $ 0.7 M_{\odot}$ in light of updated halo models.}
\label{fig:LISA observables for the halo}
\end{figure*}

\begin{figure*}[htbp] 
\centering
\includegraphics[width=0.75\textwidth]{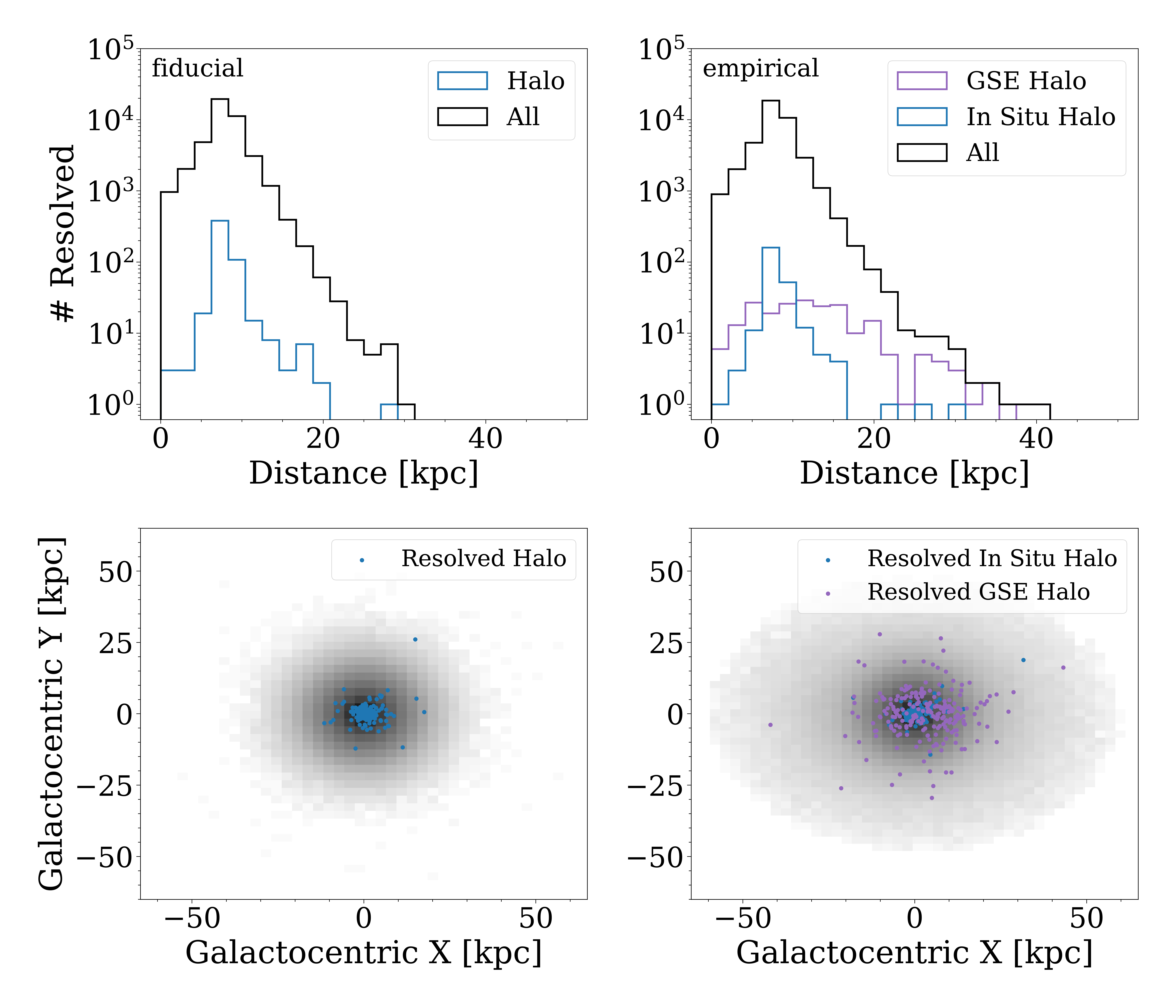}
\caption{The top panel shows the heliocentric distance of the DWDs with $SNR  >  7$. The black lines corresponds to all the other components. The bottom panel shows the Galactocentric positions of the resolved sources in the fiducial halo in blue, the GSE halo in purple and the in situ halo in blue as scatter points. Below the scatter points is the 2D distribution of the density of the DWDs in the Galaxy model from all the components. LISA will be able to resolve sources according to the GSE halo distribution and therefore see the observed halo.}
\label{fig:distances and positions}
\end{figure*}

\subsection{Milky Way shape}
\label{sec:milky-way-shape}

Our empirical model introduces new distance distributions for the halo which impact the distances of DWDs in our LISA-resolved population. The distance distribution of the entire resolved population is shown in Figure \ref{fig:distances and positions}. The top row shows the radial distance distribution for all resolved DWDs in each Galaxy model in black and highlights the halo distance distributions in color. The empirical halo is split into the spherically symmetric in-situ halo in blue and the triaxial GSE halo in purple. We find that the inclusion of the GSE halo increases the distance of the resolvable DWDs to $\sim40\,\rm{kpc}$, but that the bulk of the population is concentrated toward the Galactic center at $8\,\rm{kpc}.$

Thus, DWDs observed beyond $\sim\rm{30}\,kpc$ likely originate in the GSE and should be dominated by He-hosting DWD binaries with low chirp masses. The bottom panel of Figure \ref{fig:distances and positions} shows the resolved halo LISA sources as scatter points in blue and purple as in the top panel and a two-dimensional histogram of the DWD population radiating GWs in the LISA frequency band underlaid in grey.
In our empirical model, the resolved LISA sources are more prolate and
less concentrated in the Galactic center than the LISA-resolved sources in the fiducial halo. This suggests that LISA may be able to differentiate between spherical and triaxial halo populations. We leave a detailed investigation of this possibility to future work.

\section{Comparison to previous works} \label{sec:cite}

Several previous population synthesis studies have investigated the contribution to the LISA signal from DWDs residing in the Galactic halo and thick disk. We discuss each of these below in the context of this work.

\citet{benacquista2006} studied the effects of differing scale height assumptions of the Galactic disk on the LISA confusion noise signal. They find that when keeping the number of DWDs in the disk fixed, but varying the scale heights and thus the average distance of the population, there is not a strong effect on the height of the confusion-limited foreground. However, since the local space density decreases with increasing scale height, a larger scale height will necessarily result in a smaller amount of nearby bright sources that also are at an increased distance, making them less easily resolvable. While our study keeps the spatial distribution of the thick disk constant between the fiducial and empirical models, we do change the distribution of the halo, effectively decreasing the space density by increasing the average distance to halo DWDs. Similar to \citep{benacquista2006}, we find that this modification does not significantly alter the strength of the foreground, but can lead to a slight reduction in the number of sources that are resolvable.

\cite{ruiter2009contribution} investigated the contribution of the halo to the LISA signal from DWDs, in comparison to the disk and bulge, using assumptions similar to our fiducial model and the \texttt{StarTrack} population synthesis code. They assume a bursty star formation period a $t = 0 \, \rm{Gyr} $ for the halo and a maximum age of $13\,\rm{Gyr}$, whereas we assumed an age of $14 \, \rm Gyr$ with a 1 Gyr burst of star formation. They use a spherically symmetric model and a low metallicity matching our metallicity of $ \rm [Fe/H] = -2.3 $, which we also adopt in this work as our fiducial model. As in our fiducial Galaxy model, the disk and bulge are assigned a solar metallicty and an age of $10 \, \rm{Gyr}$. In the model of the bulge, they assume a $1 \, \rm{Gyr}$ bust of formation while the disk has a constant star formation rate within those $10 \, \rm{Gyr}$.  However, they do not consider a separate thick disk. 

Rather than changing the spatial distribution of halo DWDs, they change the relative number of DWDs in the halo by modifying the total mass of the halo made up of DWDs. They find that increasing the relative fraction of halo mass made up by DWDs increases both the number of DWDs that contribute to the foreground and that are resolved. Similar to this study, they also find that the disk and bulge components of the Galaxy dominate the DWD population that LISA will observe.

Other LISA forecasts use a different set of assumptions as their star formation history. \cite{Lamberts_2019, thiele2023applying, tang2024predictinggravitationalwavesignals} use a cosmological zoom-in simulation of a galaxy of Milky Way mass from the 'Latte' suite of the FIRE-2 simulations, considering star particles to be within 300 kpc of the Galaxy. This differs from our approach because their models do not separately consider each Milky Way component and their corresponding metallicity, age and spatial distribution. 

We recover a larger percentage of resolved sources in our models than \citet{Lamberts_2019}. This is likely because they use an iterative process that removes binaries with $\rm SNR > 7$ relative to a median-smoothed power spectrum of the LISA noise signal until the foreground shape and detection catalog converge. In this work, we apply an analytic SNR approximation using \texttt{LEGWORK} which can overestimate the number of resolved sources.

\cite{tang2024predictinggravitationalwavesignals} use \texttt{\texttt{BPASS}} to generate a binary population and \texttt{TUI} to evolve the DWDs based on gravitational emission. They then use \texttt{LEGWORK} and 
\texttt{PHENOMENA} to calculate the SNR of the sources, finding a reduced number of DWDs relative to \citet{Lamberts_2019}, \citet{thiele2023applying} and this work. 
This is likely due to the different treatment of mass transfer and CE as discussed in \citet{vanzeist2025}.

\section{CONCLUSION} \label{sec:conclusion}

In this paper, we simulate the Galactic population of DWDs using different star formation history assumptions for the halo and thick disk. Our fiducial model is inspired by previous population synthesis predictions while our empirical model includes data-driven assumptions. Our empirical model includes both an in-situ, spherically symmetric halo component as well an accreted GSE halo component following \cite{han2022} and an updated thick disk metallicity.

Our updates to the thick disk and halo do not result in a significant change to the LISA DWD foreground. However, we find that an increased metallicity for the thick disk and incorporating the younger, higher-metallicity GSE component leads to a reduction in the formation of DWDs. 

We also find that updating the star formation history of the Galaxy changes features in the observable properties of resolved DWDs. In particular, we predict that a higher metallicity thick disk will result in more COCO DWDs with chirp masses near $ 0.6 \, \rm M_{\odot} $. Similarly, incorporating the younger GSE component reduces the number of COCO DWDs and eliminates all ONeX DWDs in the halo.

Lastly, we find that including the triaxial GSE halo component modifies the distance distribution of resolved DWDs, extending the maximum distance to $40\,\rm{kpc}$. The triaxiality of the halo is also observable in the 3D positions of DWDs in the LISA band. In particular, DWDs with distances beyond $ 30 \, \rm{kpc}$ predominantly reside in the GSE component. Future observations which refine our understanding of the Milky Way's formation history will strengthen forecasts of stellar-origin binaries observable by LISA.

\begin{acknowledgments}
    A-M.A., K.B acknowledge support from NASA LISA Preparatory Science Program Grant Nos. 80NSSC24K0361 and P80NSSC26K0339 and thank the \texttt{COSMIC} research group at Carnegie Mellon University for feedback and guidance. K.B. further acknowledges support from the Falco-DeBenedetti Early Career Professorship and helpful discussions with Daniel Horta Darrington.
\end{acknowledgments}

%


\software{
This work made use of the following software packages: \texttt{astropy} \citep{astropy:2013,astropy:2018,astropy:2022,astropy_15473616}, \texttt{Jupyter} \citep{2007CSE.....9c..21P,kluyver2016jupyter}, \texttt{matplotlib} \citep{Hunter:2007}, \texttt{numpy} \citep{numpy}, \texttt{pandas} \citep{mckinney-proc-scipy-2010,pandas_17229934}, \texttt{python} \citep{python}, \texttt{COSMIC} \citep{Breivik2020,COSMIC_20030671}, \texttt{Cython} \citep{cython:2011}, \texttt{h5py} \citep{collette_python_hdf5_2014,h5py_7560547}, \texttt{legwork} \citep{LEGWORK_joss,LEGWORK_apjs,legwork_15579534}, \texttt{schwimmbad} \citep{schwimmbad}, and \texttt{seaborn} \citep{Waskom2021}.
Software citation information aggregated using \texttt{\href{https://www.tomwagg.com/software-citation-station/}{The Software Citation Station}} \citep{software-citation-station-paper,software-citation-station-zenodo}.
}

\bibliography{LISAMW.bib}{}
\bibliographystyle{aasjournal}



\end{document}